\documentclass[twocolumn,prl,aps,showpacs,superscriptaddress]{revtex4}
\usepackage{xspace,amsmath,amsfonts,amsthm,amssymb,amsbsy,graphicx,color}

\begin{document} 

\newtheorem{theo}{Theorem} 
\newtheorem{lemma}{Lemma} 

\title{Engineering Superposition States and Tailored Probes for Nano-resonators Via Open-Loop Control} 

\author{Kurt Jacobs}
\affiliation{Department of Physics, University of Massachusetts at Boston, 100 Morrissey Blvd, Boston, MA 02125, USA}
\affiliation{Hearne Institute for Theoretical Physics, Louisiana State University, Baton Rouge, LA 70803, USA}

\author{Lin Tian} 
\affiliation{Department of Applied Physics and E. L. Ginzton Laboratory, Stanford University, Stanford, CA 94305}

\author{Justin Finn}
\affiliation{Department of Physics, University of Massachusetts at Boston, 100 Morrissey Blvd, Boston, MA 02125, USA}

\begin{abstract}
We show that a nano-resonator can be prepared in mesoscopic superposition states merely by monitoring a qubit coupled to the square of the resonator's position. This works for thermal initial states, and does not require a third-order nonlinearity. The required coupling can be generated using a simple open-loop control protocol, obtained with optimal control theory. We simulate the complete preparation process, including environmental noise. Our results indicate the power of open-loop control for state engineering and measurement in quantum nano-systems. 
\end{abstract} 

\maketitle 

The quantum behavior of nano-mechanical resonators is an area of increasing activity~\cite{LaHaye04, Zhang05, Naik06, Poggio07, Xue07}. These devices offer the exciting prospect of observing quantum behavior in a macroscopic mechanical system~\cite{Armour02, Wei06}. They can also be integrated with superconducting circuits, and have potential applications such as sensing~\cite{Li07} and information processing~\cite{Cleland04}. In addition, superconducting ``stripline'' resonators are now realizing  quantum effects previously the exclusive domain of quantum optics. Recently two major advances have been made in this area: observation of the discrete energy levels of a stripline resonator~\cite{Schuster07}, and a controllable single-photon source~\cite{Houck07}. 

The task of preparing a nano-mechanical resonator in a nonclassical state has yet to be achieved experimentally. Here we consider preparing a superposition of two spatially separated wave-packets, a primary goal in this field. For a resonator to evolve to such a state itself requires that it possess a third-order nonlinearity ($(a^\dagger a)^2$ or equivalently an $x^4$ potential)~\cite{Birula68}, and the nonlinearities that exist in nanoresonators, at least at present, are too weak for this purpose~\cite{Woolley08}. A technique presented in~\cite{Law96} allows a resonator to be prepared in an {\em arbitrary} state via an auxiliary qubit, but this is best suited to states with only a few phonons. Here we show that superposition states can be prepared using an entirely different process. This requires only a continuous measurement of a qubit that is sensitive to the square of the resonator's position. We show that this preparation process can implemented by applying open-loop control to the qubit~\footnote{The term ``open-loop control" refers to any time-dependent control inputs that are predetermined {\em before} the control takes place (i.e. control not using feedback).}. This approach was inspired by~\cite{Jacobs07x}, where it was shown that control sequences applied to an auxiliary system could be used to generate nonlinearities in a target system.  While the work in~\cite{Jacobs07x} provides a proof of principle, we suggest that a better approach is to find control protocols using ``optimal control theory" (the name given to a specific optimization technique for open-loop control protocols)~\cite{Palao02}. This has been exploited to great effect in femtosecond control of chemical interactions~\cite{Tannor85}, and more recently for constructing quantum gates and entangling operations in mesoscopic systems~\cite{Sporl07, Grace07, Galve08}. The primary question is the time-scale on which the control must operate, as this determines the feasibility of the procedure.  While we explicitly consider mechanical resonators in what follows, the techniques apply, virtually without alteration, to superconducting stripline resonators. 

We now turn to our first result, that mesoscopic superposition states can be prepared purely by monitoring the square of a resonator's position. Under such a measurement, in addition to the evolution due to the resonators Hamiltonian, the density matrix, $\rho$, evolves as~\cite{JacobsSteck06} 
\begin{eqnarray} 
  d\rho \! & = & \! -k [x^2,[x^2,\rho]] dt + \sqrt{2 k} ( \{ x^2, \rho\} - 2\langle x^2\rangle \rho ) dW  \; \; 
  \label{sme} 
\end{eqnarray} 
where $\{ x^2, \rho\} = x^2 \rho + \rho x^2$ denotes the anticommutator, $x \equiv a + a^\dagger$ is the dimensionless position operator ($a$ is the resonator annihilation operator), and $k$ is the strength of the measurement (characterizing the rate which the measurement extracts information). 

To provide a physical realization of a measurement of $x^2$ we employ a Cooper-pair box (CPB) as a probe system:  we couple the CPB to the resonator via the interaction $\mu \sigma_z x^2$ (we will consider how to engineer this coupling below), and perform a continuous measurement of $\sigma_x$ on the CPB with strength $\kappa$. This measurement is described by Eq.(\ref{sme}), with $x^2$ replaced by $\sigma_x$, and can be realized by coupling the CPB to a superconducting stripline resonator, a method that has already been demonstrated with excellent resolution~\cite{Houck07}. The analysis in~\cite{Jacobs07b} indicates that $\kappa \sim 10^{9}~\mbox{s}^{-1}$ is realistic, given an average value of $10^{5}$ photons in the stripline. Here we choose $\kappa = 8\mu = 8\omega'$, where $\omega'$ is the {\em effective} frequency of the resonator as seen by the CPB~\footnote{The effective frequency is $\omega' = \nu -  \Omega + \omega $, where $\Omega$ is the CPB frequency, and $\nu$ is the frequency at which the interaction strength is modulated~\cite{Jacobs07b}.}. In all our simulations we  choose $\omega' = 2\pi f' = 2\pi f/100$, where $f = 100~\mbox{MHz}$ is the resonator frequency. 

\begin{figure}[t] 
\leavevmode\includegraphics[width=1\hsize]{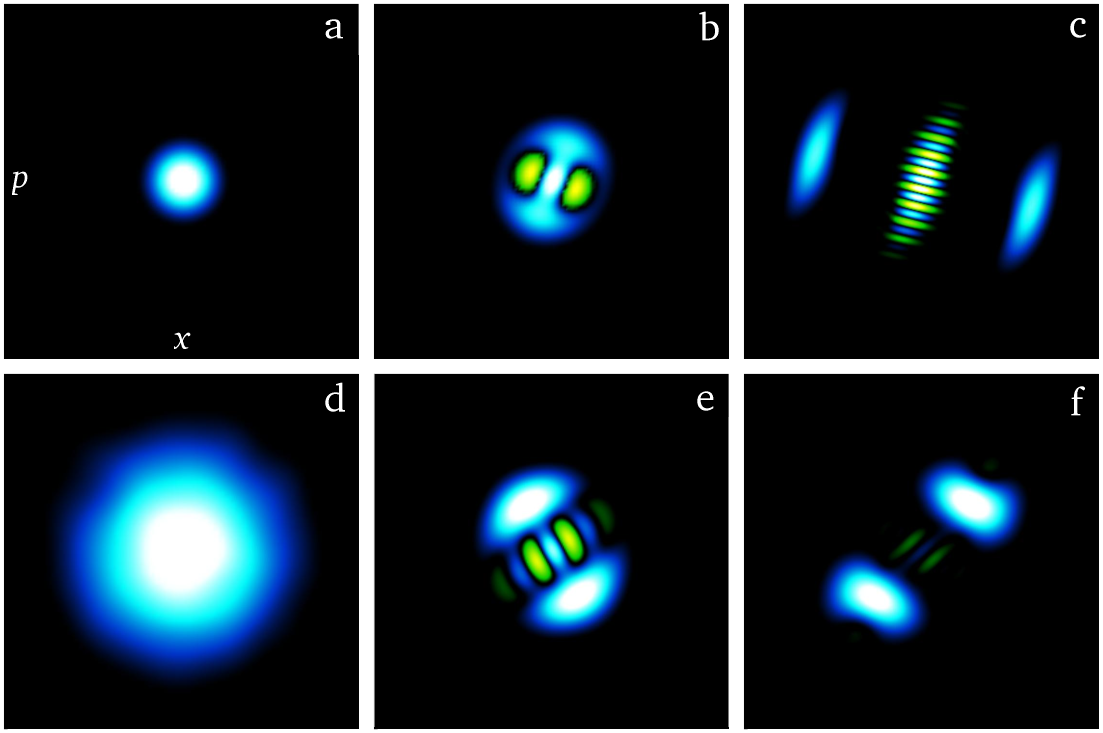} 
\caption{(Color online) Snapshots of the Wigner function for a harmonic oscillator with effective frequency $f' = 1/\tau = 1\mbox{MHz}$, subjected to a continuous measurement of the square of position. Luminosity denotes the absolute value of the Wigner function (online: blue is positive and green negative). The $x$-axis is dimensionless position, $x = a + a^\dagger$, and the $y$-axis is $p = -i(a - a^\dagger)$. Both axes cover the interval $[-7.1,7.1]$. Top row, temperature $T=0$: a) $t=0$ (vacuum state); b) $t= 6.25 \tau$; c) $t = 13.75 \tau$. Bottom row, $T=22.7~\mbox{mK}$: d) $t= 0$ (thermal state);  e) $t= 17.625 \tau$; f) $t= 18.5 \tau$.} 
\label{fig1} 
\end{figure}

For reasons that will be explained later, we implement the indirect measurement of $x^2$ by alternating between the interaction and the measurement of $\sigma_x$, with each  switched on for a duration of $1/(160f)$. We simulate this process with the resonator at  both zero temperature and $T= 22.7~\mbox{mK}$, with the resonator quality factor $Q=10^5$~\footnote{To simulate the thermal environment we use the model of Brownian motion described recently in K. Jacobs, Eprint: ArXiv:0807.4211. This is a stochastic Schr\"{o}dinger equation, and can thus be implemented using a wave-function Monte Carlo method. There is no special significance to the value $22.7~\mbox{mK}$.}. At this temperature resonator's mean phonon number is $\langle n \rangle = 4.2$~\cite{Caldiera89}. These numerically intensive simulations were performed using a regenerative Monte Carlo algorithm recently developed by one of us, running on a 128-node parallel computer~\footnote{K. Jacobs, in preparation.}. Snapshots of the Wigner function for the resonator in these two cases are shown in Fig.~\ref{fig1}. At $T=0$ a coherent superposition of two localized wave-packets is formed, and these slowly increase in separation. At $T \simeq 23~\mbox{mK}$ this superposition is still created, although it looses its coherence when the separation is sufficient for the thermal decoherence to overpower the purification due to the measurement. The coherence can be preserved for larger separations by increasing either $Q$, or $\mu$ and $k$.  

To understand the creation of superposition states, first note that a measurement of $x^2$ does not distinguish between between positive and negative $x$. Thus oscillating states whose probability densities are symmetric in position are stable under this measurement. The superposition states that are created have this form. Second, whenever a continuous measurement of $x^2$ produces a non-zero result $\tilde{x}^2$ for any appreciable time on an initial vacuum state, which it must do at some point because the stream of measurement results is random, this produces a state with peaks close to $x = \pm \tilde{x}$. This is because $\tilde{x}^2$ becomes the most likely square displacement of the resonator. The resulting symmetric superposition tends to persist because it is stable under the measurement.  
This process continues, in which new measurement results randomly change the twin peaks of the position density. The separation of the peaks thus undergoes a random walk on the real line, with the addition of a slow drift upwards. This drift is due to the fact that the measured observable does not commute with the Hamiltonian, and so feeds energy into the system. This can create superposition states with arbitrarily large separation. 

To generate superposition states using the above method, we must realize the coupling $\mu\sigma_z x^2$ between the resonator and the CPB. One can readily obtain the interaction $\sigma_z x$ merely by placing a charge on a resonator situated adjacent to the CPB. To obtain an $x^2$ interaction, we note that it was shown in~\cite{Jacobs07x} that, given the interaction $\sigma_z A$ between an auxiliary qubit and a ``target" system, an effective Hamiltonians for the target of the form $A^n$ could be generated by manipulating the qubit. The method in~\cite{Jacobs07x} can be adapted for the present problem, but it is not especially practical.  

To provide a realistic control protocol for use in a solid-state circuit, given the constraints on switching rates or other resources, we apply optimal control theory~\cite{Palao02}. In our case we want to generate evolution corresponding to the effective Hamiltonian $H_{\mbox{\scriptsize eff}} = \hbar\omega' a^\dagger a + \hbar\mu\sigma_z x^2$, given the physical Hamiltonian 
\begin{equation} 
  H_{\mbox{\scriptsize phys}} =   \hbar \omega a^\dagger a +  \hbar\Lambda(t) \sigma_z x + \Delta E \sigma_x . 
\end{equation} 
Here $\Delta E$ is the Josephson energy of the CPB, and $\Lambda$ is the strength of the resonator-CPB interaction. The CPB is operated at the degeneracy point where the qubit is largely protected from the low-frequency noise.  We first modulate the interaction strength so that $\Lambda(t) = 2\lambda \cos(\nu t)$, with $\nu = \Delta E/\hbar - (\omega - \omega')$. Moving into the interaction picture with respect to the CPB, and dropping the high frequency terms, results in the Hamiltonian 
\begin{equation}
  H =   \hbar \omega' a^\dagger a +  \hbar\lambda \sigma_z x 
  		+ H_{\mbox{\scriptsize ctrl}}(t) ,  
  \label{Hsim}
\end{equation}
where we have now added a time-dependent control Hamiltonian. We take the control Hamiltonian to be of the form $H_{\mbox{\scriptsize ctrl}} = \hbar [ c_x(t) \sigma_x + c_y(t) \sigma_y ]$. This Hamiltonian can be implemented by applying a time-dependent magnetic field to the SQUID loop of the qubit~\cite{Makhlin00}. We expect that for the control protocol to work well, the physical interaction strength should be significantly larger than $\mu$. We therefore chose (arbitrarily) $\lambda = 200\mu = 200\omega'$. With $\omega' = 2\pi\mbox{MHz}$, this means $\lambda = 4\pi\times 10^8~\mbox{s}^{-1}$, which is about a factor of 10 higher than the estimate for realistic values given in~\cite{Jacobs07b} for nano-mechanical resonators, but  realistic for coupling with superconducting resonators~\cite{Schuster07,Houck07}. We then use optimal control theory to search for a control Hamiltonian to generate $H_{\mbox{\scriptsize eff}}$ with $\mu = \omega'$. Specifically, we search for functions $c_x(t)$ and $c_y(t)$ to generate the evolution $\exp(-iH_{\mbox{\scriptsize eff}} \Delta t)$, for a time interval $\Delta t = 1/(320 f)$: this value is chosen so that $\Delta t \ll 1/f$, to realize a quasi-continuous measurement. Note that the larger $\lambda$ and $\kappa$ are, the larger is the measurement rate for $x^2$, important to beat thermal noise. 

\begin{figure}[t] 
\leavevmode\includegraphics[width=1\hsize]{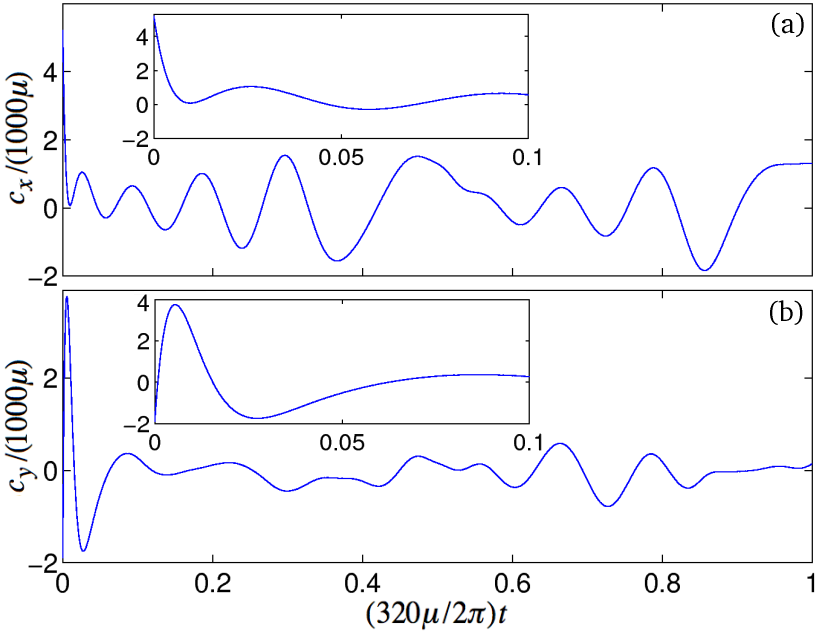} 
\caption{A control protocol for generating the effective interaction $\mu \sigma_z x^2$ between a nano-resonator and a qubit, where the physical interaction is $200\mu  \sigma_z x$, and the effective frequency of the resonator is $\omega' = \mu$. The control Hamiltonian is $H(t) =  \hbar [c_x(t) \sigma_x + c_y(t) \sigma_y]$, with $c_x$ shown in (a), and $c_y$ shown in (b). The insets give an expanded view of the first tenth of the protocol.} 
\label{fig2} 
\end{figure} 

The control Hamiltonian we have obtained is shown in Fig.~\ref{fig2}.  For the majority of the protocol, the rate at which $c_x$ and $c_y$ must be varied is less than $100 \mu $ per $\Delta t/10$. (The exception to this is in the first 40th of the protocol, where the Hamiltonian is required to change about ten times faster.) For $\mu = 2\pi \times 1 \mbox{Mhz}$, this gives a rate of change of $1 \mbox{GHz}$ per $3 \mbox{ns}$, which is readily achievable with current CPB circuits~\cite{Yamamoto03,Steffen06}. This is probably about the largest interaction strength that one can reasonably engineer with this protocol; taking $\mu = 2\pi \times 10 \mbox{MHz}$ requires a rate of $10 \mbox{GHz}$ per $0.3 \mbox{ns}$. This can be done with fast electronics, but would be restricted by the induced voltage in the SQUID loop~\footnote{Note that the limit on $\mu$ does not place any particular restriction on the frequency of the resonator, since $\omega'$ is merely the {\em effective} frequency of the resonator.}  

It turns out that the protocol presented above does not generate $H_{\mbox{\scriptsize eff}}$ with quite enough accuracy. To fix this we modify  
the protocol slightly to improve its symmetry: we apply it once, flip the sign of the resonator-CPB interaction, and then apply the protocol again with the signs of $c_x$ and $c_y$ flipped. This symmetrizes the evolution generated by $H_{\mbox{\scriptsize eff}}$ under a flip of the CPB charge states. The resulting protocol takes a time of $1/(160 f)$, plus the time to flip the sign of the interaction.    

To generate the effective $x^2$ measurement, we rapidly switch between the control protocol and a continuous measurement of the $x$-component of the CPB, at time intervals of $1/(160 f)$, just as we did previously. The reason we alternate the measurement with the control protocol is that the measurement interferes with this protocol if they are performed together. 

\begin{figure}[t] 
\leavevmode\includegraphics[width=1\hsize]{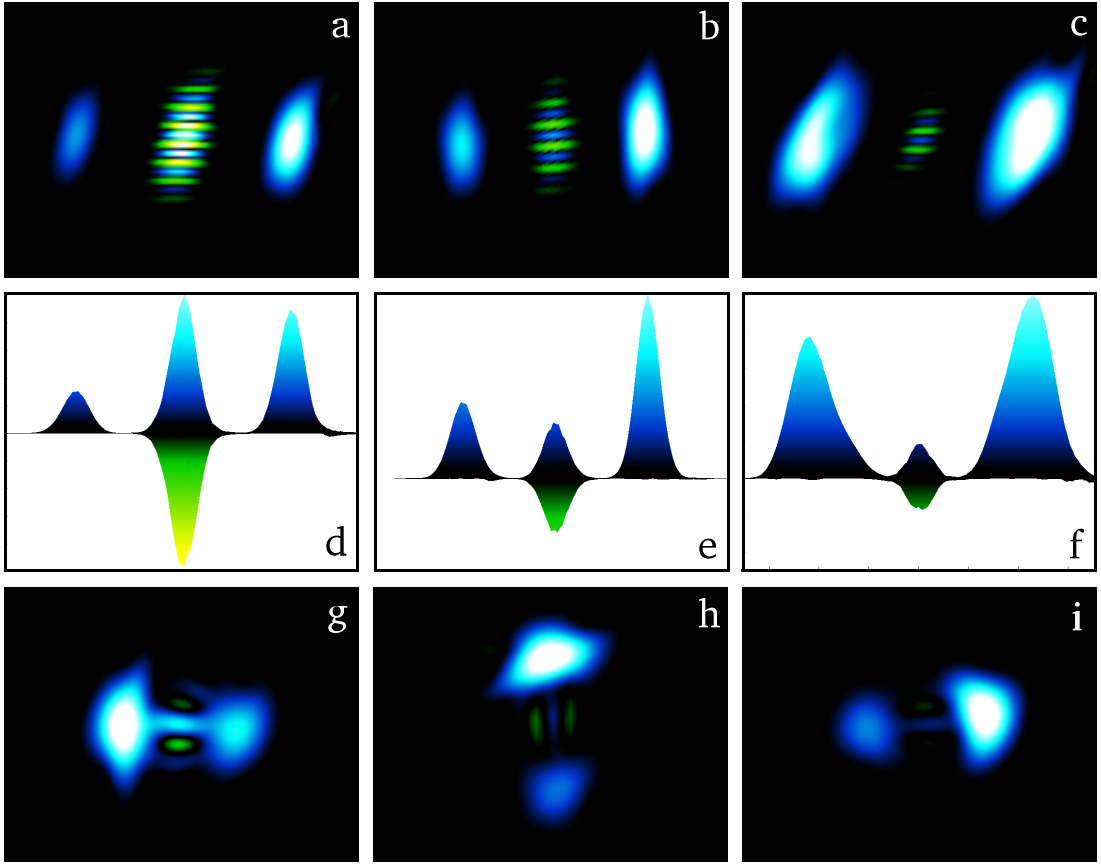} 
\caption{(Color online) The Wigner function for a harmonic oscillator measured via a Cooper-pair box (CPB). Plots (a)-(c): Mesoscopic superposition states generated by the measurement for various CPB decoherence rates, $\gamma$, and $T=0$. The phase-space scale is the same as Fig.~\ref{fig1}. (a) $\gamma= 0$; (b) $\gamma = 0.1 f'$; $\gamma = 0.5 f' $. Plots (d)-(f) are the respective side-views of plots (a)-(c). Plots (g)-(i) are a sequence where the resonator is initially prepared in the ground state, with $\gamma=0.1f' $ and the bath at $T=30 \mbox{mK}$.} 
\label{fig3} 
\end{figure}

We first simulate the entire open-loop control process, with the resonator at zero temperature, and display a snapshot of the resulting evolution of the nanoresonator in Fig.~\ref{fig3} (a). The procedure produces a superposition state as expected, although this state is no longer symmetric, (the heights of the two wave-packets are different). This asymmetry is not especially important, so long as both packets are appreciable. In Fig.~\ref{fig3}(d) we show a side view of the Wigner function, giving the heights of the interference fringes relative to the wave-packets. 

To complete our analysis we consider the combined effects of environmental dephasing on the CPB, and non-zero temperature. The decoherence rate due to this dephasing is much faster than the thermal damping rate of the resonator. To date, the best realizations of CPB's have a decoherence rate of about $10^6 \mbox{s}^{-1}$ at the degeneracy point~\cite{Vion02, Lehnert03}. It is therefore important to examine the effects of this decoherence on our control protocol. To this end we simulate the entire measurement/control process, including dephasing of the CPB qubit in all directions at rate $\gamma$. This requires adding the term  $-\gamma \sum_{i} [\sigma_i,[\sigma_i,\rho]] dt$, where $i=x,y,z$,  to the evolution of the density matrix. We choose dephasing equally in all directions to represent a worst-case scenario.  

We first simulate the evolution when the resonator is at zero temperature. The results are presented in Fig.~\ref{fig3} (b/e) and (c/f), for $\gamma = 10^5\mbox{s}^{-1}$ and $\gamma = 0.5\times10^6\mbox{s}^{-1}$, respectively, with $\mu = 2\pi \times 1 \mbox{MHz}$ as above. We see that this rate of decoherence does effect the superposition states generated by the measurement, but does not destroy the superposition completely; the interference fringes still exist, albeit significantly reduced. We then simulate the evolution when the resonator is at $30~\mbox{mK}$, with a CPB decoherence rate $\gamma = 10^5\mbox{s}^{-1}$. In this case we prepare the resonator in the ground state before we start the measurement process. A number of cooling schemes have been proposed that can potentially perform this preparation (see, e.g~\cite{NRcooling}). Superposition states can still be created in this case, although they do not last long, as is shown by the sequence of snapshots in Fig.~\ref{fig3} plots (g)-(i). These snapshots are separated by a quarter of an oscillation period. 

To summarize, we have shown that a measurement of the square of a resonators position will generate mesoscopic superposition states, even when the resonator is at finite temperature. We have also presented a method to realize this measurement, using open-loop control. While our example is probably the simplest nonlinear interaction to create in this way, our results indicate that open-loop control protocols, derived using optimal control theory, may have considerable potential for engineering a range of non-linear interactions and measurements for quantum systems. While we have used a Cooper-pair box to couple to the resonator, there are other systems that can be used for this purpose. Polar molecules, for instance, can be coupled to nano-resonators in essentially the same way~\cite{Andre06}, and suffer far less decoherence. 

The main limitation on our open-loop control protocol is the time-scale required for the control. While the protocol we have presented here is adequate for our purposes, an important open question is whether there exists a protocol for this task that allows slower control frequencies. More generally, one would like to place lower bounds on the control frequencies required for many more tasks relevant to quantum state-engineering and measurement, and this will be the subject of future work.    

{\em Acknowledgments:} This work was performed with the supercomputing facilities in the School of Science and Mathematics at UMass Boston. KJ was supported by the Army Research Office and the Intelligence Advanced Research Projects Activity. 


\end{document}